\documentclass{Interspeech2024}

\usepackage{multirow}
\usepackage{multicol}
\usepackage{caption}
\usepackage{subcaption}
\usepackage{comment}
\interspeechcameraready

\title{MSceneSpeech: A Multi-Scene Speech Dataset For Expressive Speech Synthesis}

\name{Qian Yang$^{1,*}$, Jialong Zuo$^{1,*}$, Zhe Su$^{1,*}$, Ziyue Jiang$^1$, Mingze Li$^1$ , Zhou Zhao$^1$$^{\dagger}$, 
\\
Feiyang Chen$^2$, Zhefeng Wang$^2$, Baoxing Huai$^2$}
\address{
  $^1$Zhejiang University,
  $^2$Huawei Cloud,
  }
\email{qyang1021, jialongzuo, suzhe, ziyuejiang, limingze, zhaozhou @zju.edu.cn, \\ chenfeiyang2, wangzhefeng, huaibaoxing@huawei.com}


\begin{document}

\maketitle

\begin{abstract}
    
    % 介绍数据集
    % 介绍baseline

    We introduce an open source high-quality Mandarin TTS dataset MSceneSpeech (\textbf{M}ultiple \textbf{Scene} \textbf{Speech} Dataset), which is intended to provide resources for expressive speech synthesis. MSceneSpeech comprises numerous audio recordings and texts performed and recorded according to daily life scenarios. Each scenario includes multiple speakers and a diverse range of prosodic styles, making it suitable for speech synthesis that entails multi-speaker style and prosody modeling.
    We have established a robust baseline, through the prompting mechanism, that can effectively synthesize speech characterized by both user-specific timbre and scene-specific prosody with arbitrary text input. The open source MSceneSpeech Dataset and audio samples of our baseline are available at \url{https://speechai-demo.github.io/MSceneSpeech/}.
    
    %Current adaptive text-to-speech (TTS) can synthesize high-quality voice for any user. However, transferring this tailored voice to different real-life scenarios with various prosody is still a great challenge. To address this, we propose SceneAdapter, an adaptive TTS framework that leverages reference prosody speech and prompting mechanism to model scene-specific prosody and generate multiple prosody diverse speech. Along with the model, we have curated a multi-scene dataset (MSVoice) featuring prosody-rich recordings in multiple real-life scenarios. We first pre-train our model in a 400 hours large mixed bilingual dataset with Chinese and English, employing a masked prediction mechanism to model basic timbre and prosody representations. After that, we fine-tune our model on the proposed multi-scene dataset to transfer scene-relevant prosody. Experimental results indicate our model can effectively synthesize speech characterized by both user-specific timbre and scene-specific prosody with arbitrary text input. Audio samples are available at \url{https://lmzjms.github.io/SceneAdapter/}.
\end{abstract}
\renewcommand{\thefootnote}{\fnsymbol{footnote}}
\footnotetext{* Equal contribution.}
\footnotetext{$^{\dagger}$ Corresponding to Zhou Zhao~(zhaozhou@zju.edu.cn).}
\vspace{-0.3cm}

\section{Introduction}
\label{sec:intro}

Speech synthesis is designed to empower machines with the capability to produce voices that are indistinguishable from those of humans, and it is an indispensable component of AGI (Artificial General Intelligence) human-machine interaction. In the current speech synthesis domain~\cite{lajszczak2024base,jiang2023boosting,le2024voicebox,shen2023naturalspeech,ren2020fastspeech,valle-e}, the generated speech is no longer satisfied with merely synthesizing low-noise, high-definition outputs, instead, there is a shift of focus towards generating speech that is more natural, rhythmic, and expressive.

Contemporary high-expressive speech synthesis~\cite{lajszczak2024base, jiang2023boosting} implicitly learns prosodic modeling through large volumes of data, attempting to expand the prosodic distribution by extending the distribution of data collected in the wild. Nonetheless, the quality of in-the-wild data is uneven, with complex and varied distribution patterns that do not necessarily align with the prosodic distribution found in everyday conversation. Furthermore, the absence of prosodic labeling makes it challenging to exert control over the prosody.

%Contemporary high-expressive speech synthesis~\cite{lajszczak2024base, jiang2023boosting} regards prosody as a crucial attribute. It can be modeled in the fashion of Large Language Models (LLMs) through extensive data, while timbre is typically separated during the modeling process. Ultimately, it is integrated into the model during synthesis by reference-speaker. Although current models are enriching the distribution of prosodic modeling with a wealth of open-source datasets and some in-the-wild data, these datasets almost invariably lack scene labels or other tags that could classify prosody. While the data is broad-ranging and authentic, it fails to provide the conditions necessary for generating high-quality and distinctive prosody. 

Consequently, we have recorded MSceneSpeech, a high-quality Mandarin TTS dataset meticulously organized by speaker identities and scene labels, expertly crafted to enhance the modeling and transformation of prosody. MSceneSpeech contains around 15 hours of high-quality audio recorded by professionals. The scripts for each scenario have been thoughtfully curated to align with the contextual backdrop, and the accompanying audio recordings are delivered with expressive nuance, transcending mere recitation. The dataset encompasses a diverse assortment of four scene categories: Chat, News, QA (Question and Answer), and Storytelling, with each category showcasing a range of distinct speakers. Leveraging the MSceneSpeech dataset for fine-tuning enables synthesis models to adeptly execute style-related tasks, such as multi-style speech synthesis, and cross-speaker style transfer.

In the realms of style transfer and multi-style speech synthesis, numerous exceptional studies have emerged, demonstrating the capability to generate audio of superior quality across an array of stylistic nuances~\cite{jiang2023boosting,le2024voicebox,wang2018style,ren2022prosospeech,zaidi2021daft}. However, when it comes to cross-speaker style transfer, there are scant examples of work that proficiently handles the dual challenges of style transfer and voice adaptation.
%This task is of great significance. At a time when personalization and customization in speech synthesis are rapidly advancing, the ability to synthesize specific timbres as per user requirements, and to generate multiple styles based on the scene, is undoubtedly of paramount importance. 
Thus, we have also proposed a robust baseline that is capable of generating multiple styles while performing voice adaptation. Both style transfer and voice adaptation are based on arbitrary reference audio, making the process extremely convenient and unconstrained. We utilize a timbre reference speaker module, supplemented by a prompt-based style transfer module, which explicitly extracts style information from the reference audio of the desired style and integrates it into the audio generation module. Our main contributions can be summarized as follows:

\begin{itemize}

\item We release MSceneSpeech dataset that comprises diverse real-life scenario recordings, addressing the limitation of existing datasets to exert control over the prosody.

\item We present a robust baseline, with a prompt-based prosody module and a reference timbre module, able to perform voice adaptation with different prosodic patterns across various real-life scenes. 
%Such as storytelling, news broadcasting et al.
%It first extracts prosody features as duration, pitch and energy, then utilizes a masked training mechanism and Conformer-based predictors for effective prosody transfer.

\end{itemize}

%We first pre-train our model on a large-scale mixed speech dataset, then fine-tune it and conduct experiments on proposed MSVoice Dataset. The results show that SceneAdapter is competitive with existing adaptive TTS models in voice adaptation while having extra ability of prosody transfer. We also carry out comprehensive ablation studies to substantiate the effectiveness of each design in prosody transfer.

\section{Related work}
\subsection{Expressive Speech Synthesis Corpus}
%Expressive Speech Synthesis endeavors to generate natural voice with both high quality and diversity which is crucial for real-world human-computer interaction scenarios. Previous speech synthesis systems~\cite{tacotron2,ren2020fastspeech,vits} primarily focused on audio quality and intelligibility, capable of producing high-quality speech on limited-speaker datasets, and some even achieved human-level quality~\cite{naturalspeech}. However, they struggle to generate diverse speech with different speaker identities, prosodies, and styles that expressive TTS seeks to encompass. Thus, some recent works~\cite{valle-e,shen2023naturalspeech,le2024voicebox} develop large scale speech generative models that are trained on in-the-wild at the scale of tens thousands of hours with minimal supervision, which leads to much better generalization for synthesizing unseen speech styles in a zero-shot fashion.
Over the past few years, many TTS datasets such as VCTK~\cite{yamagishi2019vctk}, LibriTTS~\cite{zen2019libritts}, AISHELL~3~\cite{shi2020aishell}, and DiDiSpeech~\cite{guo2021didispeech} have been released, making a significant contribution to speech synthesis tasks. The aforementioned datasets primarily consist of reading-style data, which, despite promoting the research of high-quality speech synthesis, exhibit limitations in terms of style coverage and prosody diversity. Therefore, some prosody-rich speech datasets are constructed for expressive speech tasks, covering a variety of speaking styles and domains. For emotion, ESD~\cite{esd} and EmoV-DB~\cite{emov-db} are designed for voice adaptation tasks and emotional TTS tasks. 
%ESD~\cite{esd} is the parallel multi-lingual and multi speaker emotional speech dataset designed for voice adaptation tasks and contains five emotional classes in each language. EmoV-DB~\cite{emov-db} is designed for emotional TTS tasks. 
For style, ST TTS~\cite{styletag} presents a style-tagged TTS dataset utilizing a short phrase or word representing the style of an utterance.
%such as emotion, intention, and tone of voice. 
Several recent studies~\cite{guo2023prompttts,yang2023instructtts,ji2023textrolspeech} are proposed to control speech style through natural language prompts which are manually annotated or conducted by automatic description creation pipeline based on speech attribute labeler and large language models. Unfortunately, these datasets are either commercially unavailable (not open source) or rely on style descriptions labeled based on natural language. In light of this, we release MSceneSpeech dataset, organized by speaker identities and scene labels, aiming to provide an open-source data resource encompassing diverse real-life scenario recordings.

\subsection{Style Transfer in Text-to-Speech}
Style transfer has been studied for decades in the TTS community, with the aim of transferring the style (e.g., prosody and emotion) from a reference utterance to the synthesized target speech. GST-Tacotron~\cite{wang2018style} introduces global style tokens to learn linguistic-agnostic prosodic features for various style control and transfer. Some works~\cite{sun2020generating,sun2020fully} further study a way to include a hierarchical, fine-grained prosody representation. 
%Recently, Adaspeech4~\cite{adaspeech4} extract speaker representation by weighted combining of basis vectors through attention, which can ensure good generalization on new speakers in zero-shot scenarios. 
GenerSpeech~\cite{huang2022generspeech} proposes a multi-level style adaptor to efficiently model a large range of style conditions. Daft-Exprt~\cite{zaidi2021daft} and MegaTTS~\cite{jiang2023boosting} introduce corresponding mechanisms to tackle the cross-speaker style transfer task. The former utilizes a gradient reversal layer to penalize the prosody encoder 
%if its output contains information about the speaker identity from the reference utterance
, while the latter 
%leverages a multi-scale timbre encoder to enhance speaker adaptation and 
adopts a latent code language model to fit the distribution of prosody. However, disentangled style control is rarely considered in existing methods, which means independently controlling timbre and other style attributes such as prosody. MSceneSpeech, on the other hand, presents a promising baseline model that transfers only the vocal style from a reference speaker prompt and generates prosodic audio samples consistent with the scene reference prompt.

\section{MSceneSpeech Dataset}
%yangqian: \subsection{Data recording and annotation}
    %录制涉及到的有：“演绎”，检查音质录制环境，与场景匹配度
    %录制参考“采集合同”
    %标注是提供了场景有关文本，但由于录制人员会念错，所以会用whisper纠错，之后还会人工纠错标注
    %ethics也是会稍微提一下
%suzhe: \subsection{Data pre-processing}
    %主要涉及切割部分了，这里涉及到duration distribution，最好用数字提一下；然后我记得也是会exclude一部分数据的，在切割的时候什么标准exclude了也提一下；
%suzhe: \subsection{Data Statistics}
    %照着之前的statistics再统计一下，放在GitHub里面的prosody有关属性的统计拿到论文里；
    %Details of train/validation (development)/test splits；这个得分一下并说明一下
In this section, we present MSceneSpeech, a novel high-quality monolingual Mandarin speech dataset that encompasses a diverse range of daily scenarios. Containing rich prosody, it serves as an important resource for synthesizing expressive audio under various scenarios.

\subsection{Data recording and annotation}
Before recording the audio, we meticulously selected scenes and carefully chose texts that were highly pertinent to each respective scenario. These were then provided to professional recording artists who, based on the content of the texts and scenario, delivered the content through interpretative performances rather than mere recitation. This approach was taken to achieve a better prosodic effect. 
%Given that the same professional individuals recorded different scenarios and that multiple professional individuals recorded the same type of scenario, 
During the recording process, we required a clear distinction in the recording style for the same person across different scenarios, while the performance method for different individuals within the same style should be consistent. 
%Besides, Our recording environment was quiet, free from noise, and with background sound kept to a minimal level.

After the audio recording was completed, we conducted verification of the text annotations. Although the recording artists were instructed to pronounce according to the annotations during recording, there were still instances where the text was read incorrectly or additional paralinguistic elements were added. To address this, we initially utilized Whisper~\cite{radford2023robust} to transcribe the recorded audio. Then, if there was a significant discrepancy between the transcribed text and the annotated text, we proceeded to manually compare and correct the text.

\subsection{Data Processing}
In our data processing workflow, we controlled the duration of audio clips to fall within the 5-10 seconds range, optimizing them for ease of use. 
Our criteria for splitting audio were initially based on the presence of traditional stopping indicators, including periods, question marks, and other punctuation marks that can be considered as a stop of a sentence. 
%Audio was primarily split at periods if the resulting segments were within the 5-10 seconds window.
The audio was primarily segmented at periods, provided that the resulting clips fell within a duration of 5 to 10 seconds.
In the absence of these stopping signs, commas and other punctuation were considered for splitting under the same time constraints.
If neither punctuation was suitable, we opted to split at the nearest available punctuation. 
%As a result, a small amount of audio was longer than 10 seconds. 
%It is crucial to note that, to preserve the integrity of speaker quotations, these signs were not considered as a limiter.

After splitting, due to potential inaccuracies in alignment, we implemented Automatic Speech Recognition (ASR) proofreading, removing any segments where text similarity fell below 80\%. This filter led to the exclusion of approximately 5\% of the audio files. We further made refinements manually, removing a small amount of sentences with alignment issues. 

After curating the dataset, we divide it into training and testing subsets. The testing subset is created by selecting one speaker and their corresponding recordings from each scene, while the remaining data are allocated to the training set.

% ----------- Data statistics (provide stats for EVERY SCENE)
% 0. Dataset length, sample numbers, speaker / scene statistics,
% 1. Length distribution
% 2. Speaker similarity (Do we really need that? Probably something related to prosody is better)

\subsection{Dataset Statistics}

\begin{table}
\centering
\small
\caption{\label{tab:dataset_stats}
Audio statistics in different domains of MSceneSpeech. Speed is measured in characters per minute.
}
\begin{tabular}{ccccc}
\hline
\textbf{Scene} & \textbf{Time (hrs)} & \textbf{\# of Clips} & \textbf{Speed} & \textbf{Pitch} \\
\hline
Chat & 6.94 & 2826 & 236.0 & 92.51\\
News & 2.01 & 722 & 234.1 & 111.21\\
QA   & 2.37 & 936 & 241.3 & 114.74\\
Storytelling & 3.36 & 1275 & 228.9 & 70.00\\
\hline
\end{tabular}
\end{table}

\begin{figure}[t]
    \centering
    \small
    \includegraphics[width=\linewidth]{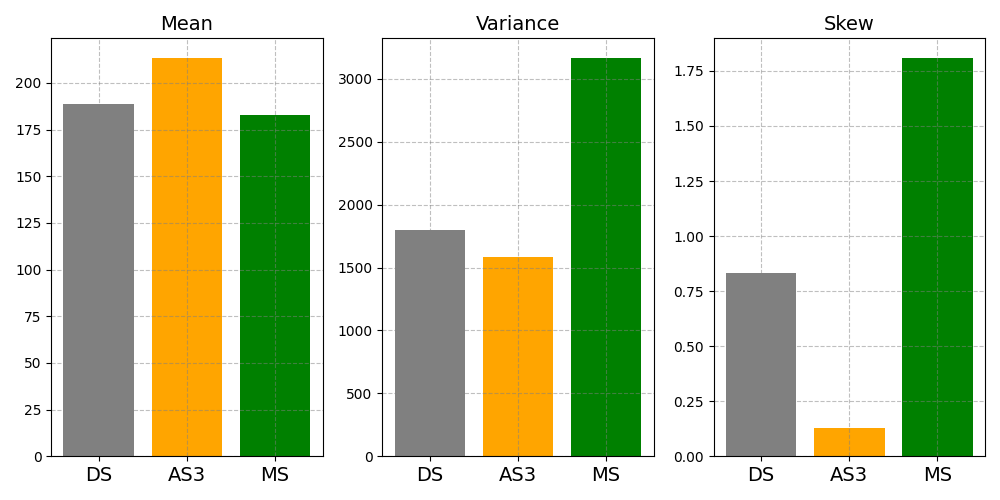}
    \caption{Comparison of Mean of pitch, Variance of pitch, and Skew of pitch across three datasets: DidiSpeech~(DS), Aishell3~(AS3), and our MSceneSpeech~(MS). The statistics represent averaged metrics from individual speakers within each dataset. Each subplot focuses on a different statistical metric.}
    \label{fig:pitch_comp}
\end{figure}

% We collect dataset statisctics and show it in Table~\ref{tab:dataset_stats}. We argue that prosody mostly exits in duration, pitch, and we can observe that these metrics varies widely across different scenes, indicating the rich prosody in this dataset. This variation presents a complex challenge for modeling both prosody and timber.
We gather statistics on the dataset and present them in Table~\ref{tab:dataset_stats}. The dataset comprises 4 distinct scenes, and 
%the total audio clips has a total length of 14.7 hours,
the aggregate duration of the audio clips amounts to a total of 14.7 hours, with a mean audio length of 9.1 seconds. 
We argue that prosodic features are predominantly found in variables such as duration and pitch, so a noticeable variation in these metrics across different scenes shows the rich prosody in our dataset. What's more, this variability introduces a challenge to model prosody and timbre separately.

Moreover, we compare our dataset with other famous multi-speaker datasets~\cite{shi2020aishell,guo2021didispeech}, focusing on the metrics of speaker mean, variance, and skew. Notably, our dataset exhibits greater pitch variance within the same speaker, indicating a richer diversity in prosody information, as shown in Fig.~\ref{fig:pitch_comp}.

\section{Proposed baseline}

%\begin{figure*}[ht]
%    \centering
%    \small
%    \subfigure[SceneAdapter]{
%        \includegraphics[width=.4\linewidth, height=.5\linewidth]{[ICASSP-2024] framwork.png}
%        \label{fig:framework}
%    }
%    \subfigure[timbre-encoder]{
%        \includegraphics[width=.3\linewidth, height=.4\linewidth]{spk_encoder.png}
%        \label{fig:spk_encoder}
%    }
%\caption{Framework}
%\label{fig:method fig}
%\end{figure*}

\label{sec:method}
\subsection{Model overview}
\begin{figure}[t]
    \centering
    \small
    \includegraphics[width=\linewidth, height=1\linewidth]{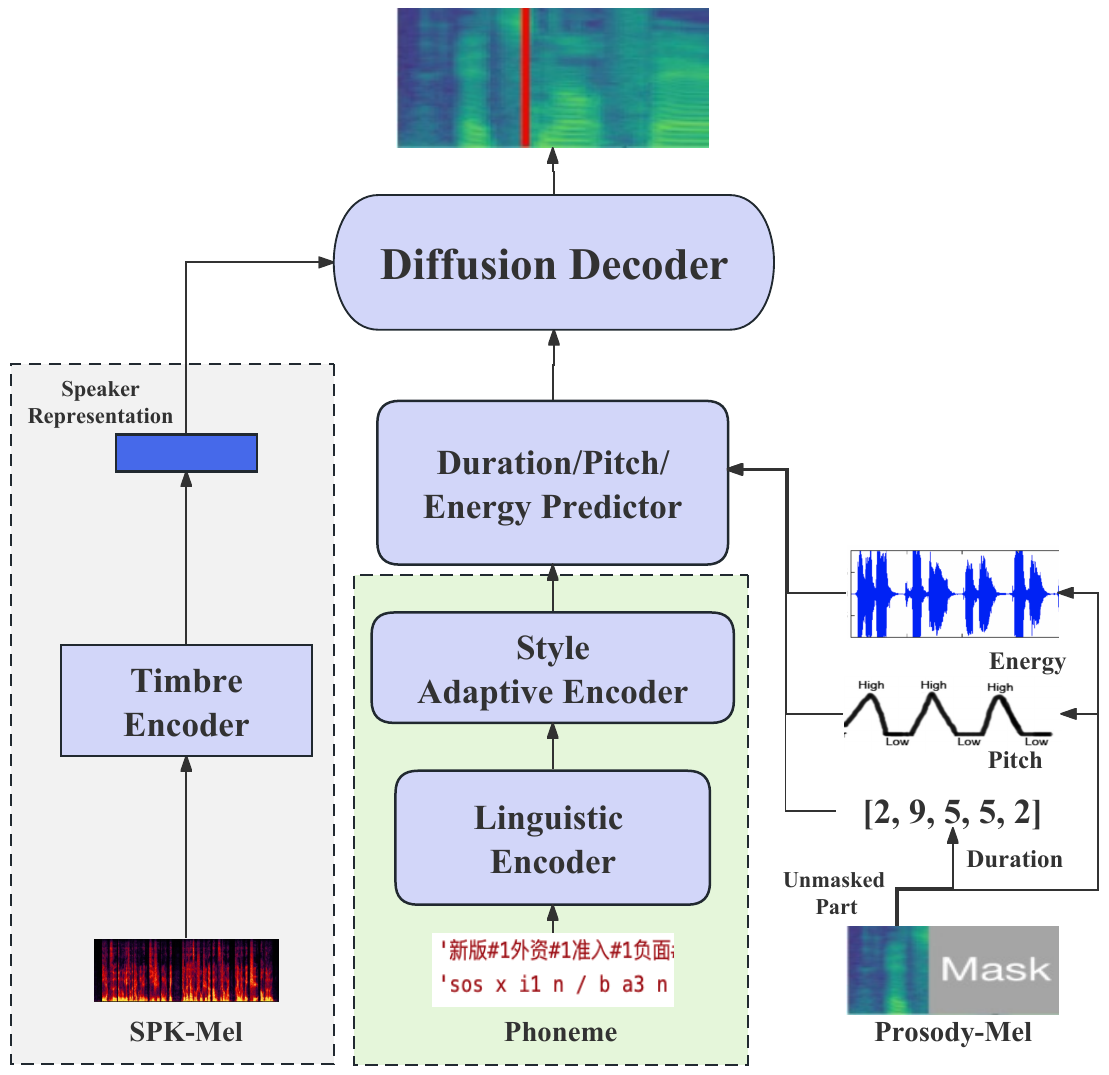}
    % \caption{The overall architecture of SceneAdapter. Duration, pitch and energy only extracted in the prompt portion, which corresponds to unmasked part in training and reference speech in inference, serve as respective condition for duration, pitch and energy predictor.}
    \caption{The overall architecture of Our Baseline. Duration, pitch, and energy are extracted from the prompt (In training:~unmasked part; In inference:~reference speech). It serves as conditions for their respective predictors. And losses are calculated only on the masked part.}
    
    \label{fig:framework}
\end{figure}

%\begin{figure}[ht]
%    \centering
%    \small
%    \includegraphics[width=\linewidth, height=.6\linewidth]{spk_enocder.pdf}
%    \caption{Framework of Timbre Encoder. Speaker Basis Vectors are initialized randomly and trained with the model.}
%    \label{fig:spk_encoder}
%\end{figure}

In this section, we introduce our proposed baseline. 
The overall architecture of our baseline is illustrated in Fig.~\ref{fig:framework}. 
The Linguistic Encoder and Style Adaptive Encoder are built upon FastSpeech2~\cite{ren2020fastspeech} and are responsible for converting raw text into a sequence of phoneme-based linguistic features.
Our Timbre Encoder is built upon Adaspeech~4~\cite{adaspeech4}, with several modifications for better adaptive voices. Specifically, a set of base vectors, which are randomly initialized and referred to as Speaker Basis Vectors, store timbre-relevant variables. We perform cross-attention between the Speaker Basis Vectors and the transformer encoder outputs of the SPK-Mel to obtain the final speaker representations. A detailed description of our Prosody-Related Module will be provided in the subsequent section.

\subsection{Prosody-Related Modules}
We argue that prosody mostly exits in duration, pitch, and energy.
It is a highly variant attribute with both local and long-term dependencies, changing rapidly over time and only has a weak correlation to text~\cite{jiang2023boosting}. 
This observation inspires us to condition prosody with a prompting mechanism.
This approach offers a stronger method for conditioning prosody compared to previous techniques that rely on implicitly extracting weaker prosodic features or employing Conditional LayerNorm~\cite{lee2021enhancing}.
% which can contribute stronger conditioning than directly adding, concatenating, or incorporating a weaker prosody condition such as Conditional Layer Normalization~\cite{lee2021enhancing}.
%Traditional: this could be deleted
% Traditional ways of conditioning encourage the model to directly add, concat or implicitly add such as Conditional Layer Normalization~\cite{lee2021enhancing}, while these ways of conditioning are proved not ideal in our experiments, resulting in poor transferring in prosody characteristics and dissimilar with prosody reference. 
% To deal with the entangled problem of timbre and prosody proposed in Section\ref{sec:intro}, we explicitly model the prosody in duration, pitch and energy, instead of implicit modeling of prosody in previous work~\cite{huang2022generspeech, ren2022prosospeech, yi2022prosodyspeech}. 
In practice, to realize the prompt mechanism, we utilize Masked Prosody Prediction (MPP), which involves masking a portion of the ground-truth duration, pitch, energy and predicting the masked values conditioning on the unmasked part. Thus, the predictors are trained to emulate the prosody presented in the reference prompt, mimicking its duration, pitch, and energy patterns. To address the issue of entangled timbre and prosody, we explicitly model prosody through its components: duration, pitch, and energy. This approach contrasts with previous work~\cite{ren2022prosospeech,huang2022generspeech,yi2022prosodyspeech} that relied on implicit modeling of prosody.
%Overall : this could also be deleted
% Overall, the duration, pitch and energy predictors in our model aim to predict corresponding part with the prompting condition of phoneme itself and the reference prosody, abandoning the timbre information existing in prosody reference.

% We explore the duration, pitch and energy predictors architecture to better learn transferring ability in prompting mechanism. We find that the original architecture in FastSpeech2 predictors, composed of pure convolutions, are unable to perform prosody transferring due to the insufficient receptive field that convolutions can offer. Thus we try to add transformer layers, finding the global information it grabs should help. The architecture of our duration, pitch and energy predictors are conformer layers~\cite{gulati2020conformer}, which contributes most in the prosody transferring processing.
% We compare different architectures of the predictors to enhance their prosody-transfer capabilities within the prompting mechanism. 
When choosing between different predictor architectures, we find the original predictor architecture in FastSpeech 2~\cite{ren2020fastspeech} performs poorly in prosody transfer, mainly because it relies solely on convolutions, providing it with a limited receptive field. To address this limitation, we employ conformer layers~\cite{gulati2020conformer} for its ability to capture global info. Such a feature has shown to be particularly effective in the process of prosody transfer in our ablation study.

\subsection{Training and inference procedures}
In this part, we outline our training and inference procedures, which primarily differ in the setting of prosody-related modules. 
% Other than that, the approach to speaker processing and decoding remains unchanged.
The speaker representations are extracted by timbre encoder, and mel-spectrograms are generated by the diffusion decoder conditioned on both the speaker and prosody representations.
% In this section, we introduce our training and inference procedures.
% which mainly differs in the prosody-related modules. Same things among training and inference are the way of speaker processing and decoding processing, in which speaker representations are all extracted as in described  by section\ref{section:timbre encoder}, and mel-spectrogram are all generated by diffusion decoder~\cite{jiang2023fluentspeech} with the condition of speaker representation and prosody representation. 
During training, we mask 60\%\footnote{In our experiments, masking 60\% of the mel-level duration, pitch, and energy ground-truth yielded the best results} posterior portion of duration, pitch, and energy ground-truths extracted from original waveform, calculating the loss for these masked portions. During inference, we concatenate the prompted duration, pitch, and energy with zeros, mirroring the masked portions during training. 

During the supervised fine-tuning stage in MSceneSpeech, we fix our Linguistic Encoder and Style Adaptive Encoder to reserve linguistic information learned from the pre-train dataset. Though we finetune other module parameters to learn rich prosody, it is crucial that the steps are controlled in a limited range so as not to break the generalization and adaptation ability trained from the pre-train dataset.

% For the prosody-related modules, we use prompting method inspired by prior masked prediction models~\cite{le2023voicebox,jiang2023fluentspeech}. In training procedure, we use masked prosody prediction (MPP) which masks a part of ground-truth duration, pitch and energy and predicts the rest of them. The preditors learn to mimic the prosody the reference prosody has offered as prompt and generate duration, pitch and energy similar with prompt. During the MPP processing we mask 60\%\footnote{60\% masks perform best in our experiment}  of Mel-level duration, pitch and energy ground-truth and calculate loss of these masked portion. Then in inference, we concat the prompt duration, pitch and energy with zero, corresponding masked part in training.

\section{Experiment}
\label{sec:typestyle}
%这部分得扩一下，把settings都说完全
%how models are initialized
%the average runtime for each model or algorithm
%The number of parameters in each model

\subsection{Experimental setup}
\label{section_exp_setup}
We pre-train our baseline on a mixed public dataset that includes both Chinese and English speech, consisting the training set of aidatatang\_200zh~\cite{aidatatang}, AISHELL-3~\cite{shi2020aishell}, DiDiSpeech~\cite{guo2021didispeech}, and VCTK~\cite{yamagishi2019vctk} datasets. This aggregated dataset comprises approximately 400 hours of speech from around 2,000 speakers. After that, we perform supervised fine-tuning on our 15-hour MSceneSpeech dataset and a few hours of internal speech corpus consisting of specific speakers.

For both pre-train and fine-tuning datasets, we convert text sequences to phoneme sequences using our internal grapheme-to-phoneme tool.
All audio files are first resampled to 16 kHz.\footnote{https://pypi.org/project/sox}
Subsequently, we extract mel-spectrograms using an FFT size of 1024, a hop size of 256, and a window size of 1024. 
The mel-spectrograms generated by our model are then converted into audio samples using HiFi-GAN~\cite{kong2020hifi}, complemented by an additional pre-trained NSF module depicted in~\cite{wang2019neural}, combining pitch information during waveform generation.

For the effective evaluation of both subjective and objective experiments, we conducted separate experiments for voice adaptation and prosody transfer. The experiments of the voice adaptation are presented in Section~\ref{exp_on_timbre}, while the experiments of prosody transfer can be found in Section~\ref{Exp_on_prosody}. Our experiments focused on the results in Mandarin, however, since our model was also pre-trained on English datasets, we have included the model's English synthesis capabilities on our demonstration page\footnote{https://speechai-demo.github.io/MSceneSpeech/}. Additionally, to test zero-shot ability of our baseline, we perform zero-shot style transfer to ESD~\cite{esd} datasets also in our demo page.

% For both public and fine-tuning datasets, we convert text sequences to phoneme sequences using our internal grapheme-to-phoneme tool. We first resample\footnote{https://pypi.org/project/sox/} all audios to 16 kHz, and then extract mel-spectrograms with a FFT size of 1024, hop size of 256, and window size of 1024 samples. The output mel-spectrograms of our model are transformed into audio samples using HiFi-GAN~\cite{kong2020hifi} with additional nsf module trained in advance.  

\subsection{Model Configuration}  
The Linguistic Encoder and Style-Adaptive Encoder are composed of multiple Feed-forward Transformer blocks~\cite{ren2019fastspeech}, with positional encoding. The dimension of the speaker embedding is set to 256. The dimension and size of the Speaker Basis Vectors are configured to 128 and 2000, respectively. The duration, pitch, and energy predictors all use the same Conformer architecture but differ in layer numbers. In the Diffusion Decoder, 
%closely aligns with the design of FluentSpeech, 
we stack 20 layers of convolution with kernel size 3 and dilation factor 1 at each layer.
% The linguistic encoder, style-adaptive encoder are consist of multiple Feed-forward Transformer blocks~\cite{ren2019fastspeech} with relative positional encoding adopted in FluentSpeech~\cite{jiang2023fluentspeech}. The dimension of speaker-embedding is 256. And the dimension and size of Speaker Basis Vectors are set to 128 and 2000. The duration, pitch and energy predictors share the same Conformer structure, but with different hyperparameters, featuring 2,3,3 layers respectively. The diffusion decoder closely resembles FluentSpeech, which we set N = 20 to stack 20 layers of convolution with the kernel size 3, and we set the dilated factor to 1 (without dilation) at each layer.

\subsection{Evaluation Methods} 
% For subjective evaluations, We conduct MOS-Q(mean opinion score on quality), MOS-S(speaker-consistency mean opinion score), MOS-P(prosody-consistency mean opinion score) evaluation on the test set to measure the audio quality via Amazon Mechanical Turk. 
For subjective evaluations, we measure three types of Mean Opinion Score (MOS): MOS-Q to assess audio quality, MOS-S for speaker consistency, and MOS-P for prosody consistency, conducted using Amazon Mechanical Turk.
% For objective evaluation, we evaluate the speaker similarly through the score of pretained Speaker Verification model wavlm-large~\cite{chen2022wavlm} and named ASV-Score. 
For objective evaluation of speaker similarity, we calculate the similarity score using the pre-trained Speaker Verification model WavLM-large~\cite{chen2022wavlm} and call it ASV-Score.
% We evaluate prosody similarity following~\cite{ren2020fastspeech}, with these steps: 1) extract pitch and energy from generated audio and reference prosody audio; 2) calculate the mean standard variation, skewness and kurtosis of the pitch and energy in each audio; 3) calculate the difference of the mean, standard variation, skewness and kurtosis between each generated audio and reference audio and average the differences among the whole set.
% To assess prosody similarity, we follow the approach outlined in~\cite{ren2020fastspeech}, which involves several steps: 1) Extract pitch and energy from both the generated and reference prosody audio. 2) Calculate the mean, standard variation, skewness, and kurtosis for pitch and energy in each audio sample. 3) Compute the differences in mean, standard variation, skewness, and kurtosis between each generated audio and its corresponding reference audio, followed by averaging these differences across the entire set.
To evaluate prosody similarity, we adopt the methodology in FastSpeech2 ~\cite{ren2020fastspeech}. This involves 1) Extracting pitch and energy metrics from generated and reference speech; 2) Calculating statistical descriptors like mean, standard variation, skewness, and kurtosis for each metric; 3) Getting the differences between generated and reference speech across the dataset, then averaging them.

\begin{table}[]
\small
\centering
\caption{Subjective evaluation on speech quality (MOS-Q) and the objective (ASV-Score) and subjective (MOS-S) evaluation on speaker consistency.}
\label{tabel:exp1}
\begin{tabular}{l|ccc}
\toprule
\bfseries Method       & \bfseries MOS-Q($\uparrow$) & \bfseries MOS-S($\uparrow$) & \bfseries ASV-Score($\uparrow$) \\
\midrule
Adaspeech 4   &   3.88 $\pm$ 0.04    &  3.85 $\pm$ 0.04    & 0.881     \\
A$^{3}$T          &   3.86 $\pm$ 0.05    &  3.76 $\pm$ 0.04   & 0.866     \\
\bfseries Our Baseline &  \bfseries 3.91 $\pm$ 0.04   &   \bfseries 4.03 $\pm$ 0.05    & \bfseries 0.884     \\ 
\bottomrule
\end{tabular}
\end{table}

\subsection{Performance on speaker consistency}
\label{exp_on_timbre}
%这一段缺少一个转折，想说一下为什么只和baseline做speaker similarity的实验（因为baseline没有做prosody和speaker同时转折的

%Our objective is to synthesize timbre and prosody both controllable speech through reference speech, and we do not ask the prosody reference speech and speaker reference coming from the same. Thus conflicting with current method, we have to perform comparisons only on timbre.
%While our model is designed for simultaneous prosody and timbre transfer, we also conduct experiments comparing its performance on speaker consistency with other Adaptive TTS systems, including A$^{3}$T~\cite{bai20223} and Adaspeech 4~\cite{adaspeech4}.
Voice adaptation is not the innovative aspect of our baseline, therefore, in the experiments, we only focus on comparing our timbre module with our backbone models, including A$^{3}$T~\cite{bai20223} and Adaspeech 4~\cite{adaspeech4}.
% We perform experiments with other Adaptive TTS systems on speaker consistency though our model is designed to perform prosody and timbre transferring simultaneously. 

% For our method SceneAdapter, we make our prosody reference the same with speaker reference, only observing the voice adaptive results. All the systems are trained firstly in mixed dataset and then fine-tuned in multi-scene dataset. 
In the experiments of our baseline, we set the prosody reference to be identical to the speaker reference, focusing solely on the results related to voice adaptation. All systems were trained using the same dataset configurations and parameter settings as described in section~\ref{section_exp_setup}.

% As shown in Table\ref{tabel:exp1}, though SceneAdapter performs unremarkable in speech quality, it outperforms the baseline method in speaker consistency, proving the voice adaptive ability apart from prosody transferring.
As shown in Table~\ref{tabel:exp1}, Our baseline showcases its capability in voice adaptation is on par with existing expert voice adaptation models, while our baseline remains its ability in style transfer.

\begin{table}[]
\small
\centering
\caption{Subjective evaluation on speech quality (MOS-Q),
prosody consistency (MOS-P) and speaker consistency (MOS-S).}
\label{tabel:exp2}
\begin{tabular}{l|cccc}
\toprule
\bfseries Method          & \bfseries MOS-Q & \bfseries MOS-P & \bfseries MOS-S \\
\midrule
\bfseries Our Baseline   &  ~0.00 &  ~0.00 &  ~0.00 \\
CNN-Predictor   &  -0.78    & -0.57      & -0.15      \\
Atten-Predictor &  -1.23     &  -0.47     &  -0.45     \\
Parallel-SPK       &  -1.10     &  -0.56     &   -0.54    \\
Addall-SPK     &  -0.26     &  -0.58     &   +0.01    \\
NS2-Prompting             &  +0.01     &  -0.06     &   -0.37    \\ 
\bottomrule
\end{tabular}
\end{table}

\begin{table}[]
\caption{Objective evaluation on speaker consistency (ASV-Score) and prosody consistency on pitch difference.}
\label{table:exp3}
\resizebox{\columnwidth}{!}{%
\begin{tabular}{l|ccccc}
\toprule
\multicolumn{1}{c|}{\multirow{2}{*}{\textbf{Method}}} & \multicolumn{1}{c}{\multirow{2}{*}{\textbf{ASV-Score}}} & \multicolumn{4}{c}{\textbf{Pitch}}                                                                                                           \\ \cline{3-6} 
\multicolumn{1}{c|}{}                                 & \multicolumn{1}{c}{}                                    & \multicolumn{1}{c}{\textbf{Mean}} & \multicolumn{1}{c}{\textbf{Std}} & \multicolumn{1}{c}{\textbf{Skew}} & \multicolumn{1}{c}{\textbf{Kurt}} \\ \hline
\midrule
\textbf{Our Baseline}                                & 0.833                                                   & \textbf{11.48}                    & 7.31                             & \textbf{0.53}                     & \textbf{3.64}                     \\
CNN-Predictor                                         & 0.831                                                   & 22.80                             & 27.23                            & 0.73                              & 5.10                              \\
Atten-Predictor                                       & 0.739                                                   & 66.50                             & 21.52                            & 2.23                              & 5.58                              \\
Parallel-SPK                                          & 0.774                                                   & 27.55                             & 16.85                            & 0.88                              & 5.10                              \\
Addall-SPK                                            & \textbf{0.845}                                          & 11.55                             & 9.18                             & 0.68                              & 4.98                              \\
NS2-Prompting                                         & 0.756                                                   & 11.50                             & \textbf{4.81}                    & 0.82                              & 5.67  \\
\bottomrule
\end{tabular}%
}
\end{table}

\subsection{Ablation Study and Prosody Transfer Analysis}
\label{Exp_on_prosody}
%这里感觉介绍用了什么模型篇幅过长，如何去删？
% We show our prosody transferring ability in ablation study, 
% comparing with the architecture which we try along the road.
We further conduct ablation studies to verify the effectiveness of each carefully designed module for prosody transfer.
%We do overall experiments on audio quality, 
%speaker consistency with speaker reference, prosody consistency with prosody reference in subjective and objective metrics. 
% We mainly consider three aspects: 1) Duration, pitch and energy predictor architecture, based on original CNN Predictor(Marked CNN-Predictor), or CNN mixed single Attention Predictor(Marked Atten-Predictor). 2) the way of speaker representation being added to the model, no shuffling speaker mentioned in Section\ref{section:timbre encoder}(Marked Parallel-SPK) or adding speaker representations on all part of the model(Marked Addall-SPK). 3) implicit modeling prosody based prompting mechanism as mentioned in~\cite{shen2023naturalspeech} (Marked NS2-Prompting). 
We focus on three key design choices: 1) Predictor architecture—either original CNN-based (CNN-Predictor) or CNN with single attention (Atten-Predictor); 2) Speaker representation—either no shuffling speaker approach (Parallel-SPK) or incorporating speaker embedding throughout all parts of the model (Addall-SPK); 3) Prosody modeling—using an implicit prompting mechanism as per NaturalSpeech2~\cite{shen2023naturalspeech} 
 (NS2-Prompting).

%We do ablation on three aspects: 1) duration, pitch and energy predictor architecture, including original CNN-based Predictor~\cite{ren2020fastspeech} and CNN blending singe Attention based Predictor. 2) the way speaker representation adding to the model. We test No shuffling speaker mentioned in Section\ref{section:timbre encoder}(Marked Parallel-SPK) and adding speaker representations on all part of the model(Marked Addall-SPK). 3) implicit modeling of prosody, which means we do not extract pitch, duration and energy explicitly and model prosody using a speech prompt encoder(Marked NS2-Prompting). We do this experiment as the prompting mechanism designed in NaturalSpeech2\cite{shen2023naturalspeech}.

% Limited by space, we put energy objective experiments in demo page and only remain pitch exp. We can observe in Table\ref{tabel:exp2} and Table\ref{table:exp3}: 1) CNN-Predictor though drops slightly in speaker consistency, perform bad in prosody consistency. 2) single attention layer is able to perform both timbre and prosody transferring. 3) parallel speaker training is worse than no-parallel since the training and inference mismatch. 4) adding speaker representations to all modules indeed contribute to better speaker consistency but harm the multiple prosody it generates. 5)implicit modeling meets the problem of entanglement of timbre and prosody, thus resulting poor speaker consistency with the influence of prosody reference.

We perform the subjective evaluation of speech quality, prosody consistency, and speaker consistency on Table~\ref{tabel:exp2} and objective evaluation of prosody consistency. We put energy evaluation on prosody consistency in the demo page and only show pitch on Table~\ref{table:exp3}. Key findings in subjective and objective metrics include: 1) CNN-Predictor shows poor prosody transfer with a slight drop in speaker consistency, and Atten-Predictor struggles to perform in both timbre and prosody transfer; 2) Parallel-SPK underperforms due to train-test mismatch, and Addall-SPK boosts speaker consistency but affects prosody diversity negatively; 3) NS2-Prompting entangles timbre and prosody, harming speaker consistency.

%\section{Limitations}

\section{Conclusion}
\label{sec:conclusion}
% In this paper, we propose SceneAdapter, an adaptive TTS model with the ability to generate multiple scene-relevant prosody. To model speech prosody, we explicitly represent it using duration, pitch and energy and perform transfer using prompting mechnism equipped with conformer predictors. Moreover, we contribute a multi-scene dataset with prosody-rich recordings in multiple real-life scenarios, prompting the modeling of expressive speech with rich prosody. Our experimental results demonstrated the effectiveness of speaker voice adaptiveness and multiple scene-relevant prosody transfer ability of SceneAdapter. In the future, we propose to generate speech with more expressiveness and high quality under the setting of adaptive TTS, and explore more ways on convenient prosody controlling.
In this paper, we introduce MSceneSpeech, an open source high-quality Mandarin expressive TTS dataset, featuring a diverse range of real-world scenarios, and aims to advance the synthesis of speech with richer prosody and to facilitate style transfer and synthesis within specific everyday scenarios. Furthermore, we introduce a strong baseline, capable of synthesizing audio with the corresponding timbre and prosody based on any given timbre reference audio and prosody reference audio. Experimental results show our baseline's superior performance in both speaker voice adaptability and multiple prosody transferability. 
%In the future, we plan to expand the scale of the dataset and continue to make it open source. 
%we will focus on enhancing the model's expressive capabilities and investigating powerful prosody control.

%\section{Acknowledgements}
%Acknowledgement should only be included in the camera-ready version, not in the version submitted for review.
%The 5th page is reserved exclusively for \red{acknowledgements} and  references. No other content must appear on the 5th page. Appendices, if any, must be within the first 4 pages. The acknowledgments and references may start on an earlier page, if there is space.

%\ifinterspeechfinal
%     The Interspeech 2024 organisers
%\else
%     The authors
%\fi
%would like to thank ISCA and the organising committees of past Interspeech conferences for their help and for kindly providing the previous version of this template.

\bibliographystyle{IEEEtran}
\bibliography{mybib}

\end{document}